# Radar in-Band Interference Effects on Macrocell LTE Uplink Deployments in the U.S. 3.5 GHz Band

Mo Ghorbanzadeh, Eugene Visotsky, Prakash Moorut, Weidong Yang
Nokia Solutions and Networks US LLC
Arlington Heights, USA
{mo.ghorbanzadeh, eugene.visotsky, prakash.moorut, weidong.yang}@nsn.com

Charles Clancy
Hume Center for National Security and Technology
Virginia Tech, Arlington, USA
tcc@vt.edu

*Abstract*—National Telecommunications and Information Administration (NTIA) has proposed vast exclusions zones between radar and Worldwide Interoperability for Microwave Access (WiMAX) systems which are also being considered as geographic separations between radars and 3.5 GHz Long Term Evolution (LTE) systems without investigating any changes induced by the distinct nature of LTE as opposed to WiMAX. This paper performs a detailed system-level analysis of the interference effects from shipborne radar systems into LTE systems. Even though the results reveal impacts of radar interference on LTE systems performance, they provide clear indications of conspicuously narrower exclusion zones for LTE vis-à-vis those of WiMAX and pave the way toward deploying LTE at 3.5 GHz within the coastline populous areas.

## I. INTRODUCTION

Mobile broadband networks will face a tremendous increase in data traffic volumes over the next 20 years. In order to meet this need, large amounts of spectrum will be a key prerequisite for any radio access network evolution. To satisfy this demand, Mobile Network Operators (MNOs) will need new spectrum allocations [1]. The created demand for more bandwidth far exceeds the available commercial spectrum and has spurred the Federal Communications Commission (FCC) to consider several potential measures including incentive auctioning and spectrum sharing, which is an elegant solution to utilizing sharable spectrum bands efficiently.

In spite its attractiveness, spectrum sharing is challenging as incumbent systems need to be shielded from harmful interferences from the incoming systems and vice-versa. For instance, in response to President's Council of Advisers on Science and Technology (PCAST)'s report [2] on realizing the full potential of the government-held spectrum, FCC's Notice of Proposed Rulemaking (NPRM) [3] proposed designating the 3550 - 3650 MHz range (aka the 3.5 GHz band) for mobile broadband. The National Telecommunications and Information Administration (NTIA) conducted a Fast Track Evaluation [4] to study the interference between radar and Worldwide Interoperability for Microwave Access (WiMAX) systems [5], which led to establishing exclusion zones larger than 400 km (Figure 1) before any 3.5 GHz Long Term Evolution (LTE) [6] systems are deployed. The cast exclusion zones rendered serving the United States (US)'s coastal regions infeasible. With over 55% of the US population lives within 50 miles of the shoreline [7], the inability to cater for this huge market which includes metropolises like New York City and Los Angeles ensues severe financial caveats for MNOs Thus, any efforts to judiciously decrease the exclusion zones is of great interest for MNOs, a desirable step toward realizing 3.5 GHz radar-LTE coexistence, and an inspiration for spectrum sharing of other bands in the future.

However, the aforementioned exclusion zones were developed from link budget analyses of radars and WiMAX systems [4] and may drastically change once the dynamic nature of the radar interference and details of the LTE link-level protocol, such as turbo-coding, advanced scheduling techniques and Hybrid Automatic Repeat Request (HARQ) are considered [6]. It is anticipated that LTE would become the preferred technology deployed in this new band. With this regard, analyzing radar interference effects on LTE systems, not WiMAX, can provide with relevant exclusion zones in the 3.5 GHz band.

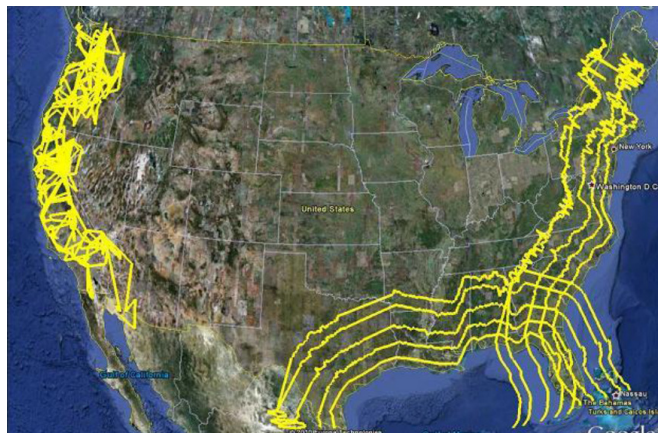

**Figure 1: NTIA set radar-WiMAX exclusion zones (area covered by yellow curves) exceeding 400 km.**

As such, this paper investigates the effects of S-band rotating shipborne radars interference into a Time Division Duplex (TDD) LTE on its uplink. The investigation relies on simulating representative radar parameters in details. In particular, rotation, antenna pattern, antenna size, pulse repetition interval (PRI), and antenna dwell time are incorporated into the simulations. Furthermore, free space path loss (FSPL) and irregular terrain model (ITM) diffraction/troposcatter loss [8] with the parameters from the NTIA's Fast Track Evaluation [4], which had led to the exclusion zones in Figure 1, were simulated to inspect the attenuation the radar signal undergo before they reach the cellular system. Besides, a 3rd Generation Partnership Project (3GPP) [9] compliant TDD LTE system level simulator was developed. The LTE simulation includes a hexagonal macro cell layout, HARQ, turbo coding, antenna pattern, moving users, indoor users, etc. Our simulations indicate certain level of LTE susceptibility to co-channel radar interference; however, the simulations imply that LTE-radar exclusion zones should be much narrower than those of the NTIA report [4]. To the best of our knowledge, no previous work has studied the effect of radar interference into LTE while fully considering the on-off characteristics of pulsed radar interference together with all the nuances of the LTE link-level protocol.

Before proceeding any further, a list of abbreviations used in the current article are written in Table 1 for convenience.

**Table 1: List of Abbreviations and their Descriptions.**

| Abbr. | Description |
|---|---|
| MNO | Mobile Network Operator |
| NTIA | National Telecommunications and Information Administration |
| FCC | Federal Communications Commission |
| ITU | International Telecommunications Union |
| IMT-A | International Mobile Telecommunications-Advanced |
| LTE | Long Term Evolution |
| UE | User Equipment |
| WiMAX | Worldwide Interoperability for Microwave Access |
| PCAST | President's Council of Advisers on Science and Technology |
| HARQ | Hybrid Automatic Repeat Request |
| TDD | Time Division Duplex |
| FSPL | Free Space Path Loss |
| ITM | Irregular Terrain Model |
| 3GPP | 3rd Generation Partnership Project |
| eNB | Evolved NodeB |
| BLER | Block Error Rate |
| APM | Area Prediction Mode |
| LoS | Line-of-Sight |
| NLoS | Non-Line-of-Sight |
| SINR | Signal-to-Interference-to-Noise-Ratio |

*A. Related Work*

The authors in [4] investigated the WiMAX-radar mutual interference and concluded large geographic separations between the two systems, precluding WiMAX deployability in the coastline. Cotton et. al. [10] performed tests using an S-band shipborne radar in San Diego littoral areas to measure temporal mean band occupancy and found that the 3.5 GHz spectrum is not often occupied by radar transmissions, thereby underlining the promising potential of the germane band for spectrum sharing. Lackpour et. al. [11] suggested a general spectrum sharing scheme based on time, space, frequency, and system-level modifications, of which the last is inconducive to real-world implementation. Khawar et. al. [12] proposed projecting a radar signal onto the null space of the interference channel to allow for spectrum sharing with futuristic radars. Ghorbanzadeh et. al. [13] developed a geographic-based spectrum sharing algorithm for radar and sectorized cellular systems with frequency reuse. However, they could not come up with any reasonable exclusion zones where their algorithms can be deployed. Ultimately, Sanders et. al. [14] performed an experiment with cabled RF connections to observe the effects of interference from radar waveforms onto a proprietary 3.5 GHz LTE evolved NodeB (eNB). In particular they considered the throughput loss and block error rate (BLER) for the uplink LTE system. Their results showed some waveforms did not have any appreciable effect on the LTE while others undermined the performance drastically. However, they did not consider any propagation models for their simulation, nor did they simulated a realistic radar system with rotation, antenna pattern, and other operational parameters (Table 2), nor did they simulate details of the LTE protocol (Table 3).

*B. Organization*

The remainder of this paper presents our simulation setup in section II, reports simulation results in section III, and concludes the paper in section IV.

## II. SIMULATION SET-UP

We simulate a pulsed radar system with representative operational parameters adopted from NTIA's Fast Track Evaluation [4] to compare exclusion zone options. Next, we leverage an LTE system level simulator to create a 7-site macrocellular system with wrap-around juxtaposed to the radar. Then, we assess the radar interference into the LTE eNBs and produce geographic separations for the coexistence.

*A. Radar Simulation*

We leveraged radar operational parameters in Table 2, adopted from NTIA [4] except for those marked with asterisks, not given due to the sensitive nature of the parameters. For those parameters, we opted to use typical values considering medium-to-large shipborne S-band radars [15]. Placing the radar 50, 100, 150, and 200 km away from the LTE system, similar to the scenario in Figure 2 (a), it affects the eNB(s) depending on the the cells radii and radiation diameter $d$ which relates to the radar-eNB distance $R$ and radar horizontal beamwith $\theta_a$ as equation (1).

$$d = 2R\tan(\theta_a) \approx 0.03R \qquad (1)$$

The radar scans 360 deg in azimuth with an angular speed 30 rpm, generating a scan time of 2 s, in which 4000 pulses (pulse repetition frequency 1/0.5 ms = 2000 Hz) of power 83 dBm, excluding the antenna gain, are emitted. The pulse-widths are 78 µs yielding in a bandwidth 13.5 kHz. Also, the horizontal beamwidth 0.81 deg creates 445 beam positions so that the antenna dwell time becomes 4.5 ms. Therefore, eNBs affected in each beam position are hit with 9 pulses as in Figure 2 (b), whose abscissa/ordinate is time/amplitude in seconds/Volts where the amplitude is the square root of the pulse power in Watts. Furthermore, the radar antenna was according to the cosine pattern [4], plotted in Figure 3, with the normalized gain as a function of the angle $\theta$ from the boresight in equation (2) where the first, second, and third expressions respectively give (i) the theoretical directivity pattern, (ii) a mask equation according to which the pattern deviates from the theoretical one at an angle corresponding to a side lobe of 14.4 dB below the main beam, and (iii) the back lobe.

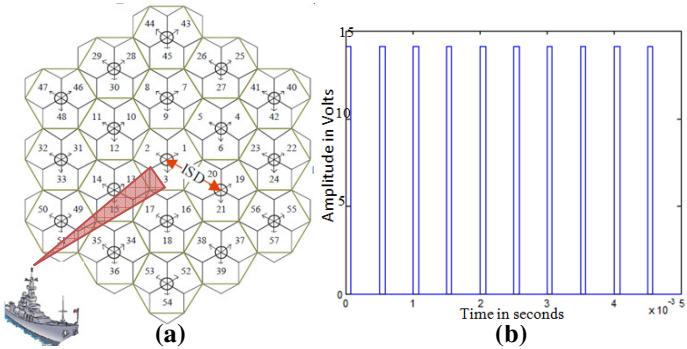

Figure 2: (a) Simulation scenario includes a shipborne radar approaching littoral zones juxtaposed to a 3.5 GHz LTE cellular network. (b) Baseband radar pulses during the antenna dwell time radiating onto the LTE system.

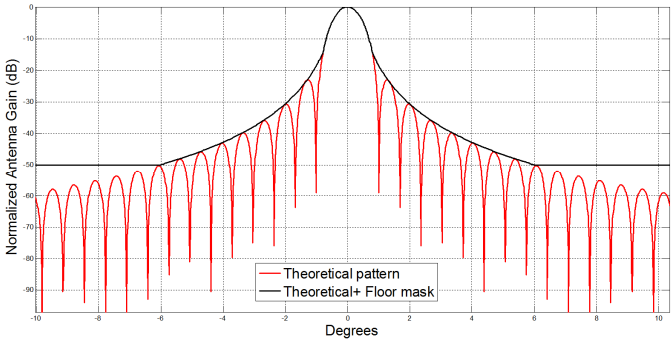

Figure 3: Radar Antenna Pattern: The black curve represents the pattern after the first side lobe is 14.4 dB below the main lobe. The back lobe is constant at -50 dB. The pattern was used in the NTIA Fast Track Evaluation [4].

$$G(\theta) = \begin{cases} \dfrac{\pi}{2} \left( \dfrac{\cos(\dfrac{68.8\pi \sin(\theta)}{\theta_{3dB}})}{(\dfrac{\pi}{2})^2 - (\dfrac{68.8\pi \sin(\theta)}{\theta_{3dB}})^2} \right) \\ -17.51 \log_e (\dfrac{2.33|\theta|}{\theta_{3dB}}) \\ -50\, dB \end{cases} \quad (2)$$

## B. LTE Simulation

The LTE system level simulation is fully compliant with the 3GPP evaluation methodology [9] and is based on the International Telecommunications Union (ITU) recommendations on International Mobile Telecommunications-Advanced (IMT-A) radio interface technologies [16]. It leverages a full-buffer traffic model and includes indoor, microcellular, and macrocellular infrastructure models. For our experiments, we considered a macro-cellular model (urban macro) with cell ISD of 500 m. All users were simulated at a pedestrian speed of 3 kmph. The network layout contains eNBs placed on a grid hexagonally (Figure 2 (a)) and the simulation included 7 sites each with 3 cells, i.e. 21 cells.

**Table 2: NTIA [4] Radar parameters, except those marked with \*, not given in [4], where we used typical parameters.**

| Parameters | Value |
| --- | --- |
| Operating Frequency | 3.5 GHz* |
| Peak Power | 83 dBm |
| Antenna Gain | 45 dBi |
| Antenna Pattern | Cosine |
| Antenna Height | 50 m |
| Insertion Loss | 2 dB |
| Pulse Repetition Interval | 0.5 ms |
| Pulse-Width | 78 µs |
| Rotation Speed | 30 rpm* |
| Azimuth Beam-Width | 0.81 deg* |
| Elevation Beam-Width | 0.81 deg* |
| Azimuth Scan | 360 deg |
| Pulse Repetition Interval | 0.5 ms |
| Distance to LTE | 50, 100, 150, 200 km |

The cells in the simulation had 120 deg sectors, and the sector antennae were directed at 90, 210, 330 deg. The LTE parameters are illustrated in Table 2, adopted from [9, 16] which include the steps to simulate an LTE system level simulator. It is worth mentioning that user equipment (UE) antennae are omnidirectional and that is why their gain is 0 dB in Table 3. The eNB antenna pattern per sector is as equation (3) in which $G_A$ and $\theta_A$ ($G_E$ and $\theta_E$) represent the antenna azimuth (elevation) pattern and angle off the antenna boresight in the azimuth (elevation) direction where $-180° \leq \theta_A \leq 180°$ ($-90° \leq \theta_E \leq 90°$), antenna azimuth (elevation) downtilt is $\theta_{A,t} = 0°$ ($\theta_E = 15°$), $A_m = 20$ dB is the maximum attenuation, and $\theta_{3dB}$ is the antenna 3dB beamwidth. The composite antenna pattern, plotted in Figure 4, can be expressed as in equation (4)

$$G_i(\theta_i) = -\min\{12(\dfrac{\theta_i - \theta_{i,t}}{\theta_{3dB}})^2, A_m\}, \ i \in \{A, E\} \quad (3)$$

$$G = -\min\{-(G_A(\theta_A) + G_E(\theta_E)), A_m\} \quad (4)$$

## C. Radar to LTE Propagaion Model

To account for the propagation loss from the radar to the eNBs, we need to consider appropriate models based on particularly the distance and terrain between the radar and LTE system. Owing to the fact that in our simulations the radar is at least 50 km away from the LTE system, the radiated electromagnetic signal sees a path loss associated with the line-of-sight (LoS) region, for which FSPL is suitable, and pursue with experiencing a diffraction loss in the non-LoS (NLoS) region, for which ITM has been the predominantly leveraged model by the FCC and NTIA as in [4]. FSPL can be expressed using equation (5) in which $f$ is the radar operating frequency in MHz, and $r$ is the distance in km at which FSPL $L_{dB,FSPL}$ in dB is requested, and $r_{LoS}$ is the border of the LoS region in km as in equation (6) [4] [4] where $y_e$ is the effective earth curvature and equals km and $h_{radar}/h_{LTE}$ is the radar/LTE antenna height as 50 m /25 m.

$$L_{dB,FSPL}(r) = 20\log(f) + 20\log(r) + 32.45, r < r_{LoS} \quad (5)$$

$$r_{LoS} = 4.1(\sqrt{h_{radar}} + \sqrt{h_{LTE}}) \quad (6)$$

**Table 3: LTE Parameters for Nokia Simulator from [15].**

| Parameters | Value |
|---|---|
| Operating Frequency | 3.5 GHz |
| Layout | Hexagonal grid |
| Mode | TDD |
| Downlink: Uplink Ratio | 2:3 |
| eNB/UE Transmit (TX) Power | 46/23 dBm |
| Macro-cell Sites | 7 |
| Indoor UE | 80% |
| Bandwidth | 20 MHz |
| eNB/UE Antenna Gain | 17/ 0 dBi |
| Inter-site Distance (ISD) | 500 m |
| Minimum UE-eNB Distance | 25 m |
| eNB Antenna Downtilt | 12 deg |
| eNB/UE Antenna Height | 25/1.5 m |
| Indoor UE | 80% |
| UE Distribution | Uniform |
| UE Mobility | 3 km/h, uniform direction |
| eNB/UE Noise Figure (NF) | 5/9 dB |
| Thermal Noise | -174 dBm/Hz |
| Service Profile | Full buffer best effort |
| UE per Cell | 10 |
| Channel Model | UMa [9] |

For instance, in our scenario, the radar antenna is 50 m high and the LTE eNBs are 25 m high. Therefore, the border for the LoS region occurs at 49.5 km. That is to say, the radar pulses undergo a propagation loss in accordance with the equation (5) for distances as far as 49.5 km away from the ship on which the radar is located. At distance beyond 49.5 km, the radar signals are within the NLoS region where diffraction loss and troposcatter loss apply to the radar signals. As for the diffraction and trposcatter losses, we leveraged the ITM in its area prediction mode (APM) [8] with the terrain roughness 10 m, LTE and radar antenna heights 25 and 50 m, ground dielectric constant 15, ground conductivity 0.005 S/m, refractivity 301 N-units, continental temperate climate, and single message mode as listed in Table 4. These parameters were adopted from the NTIA's Fast Track Evaluation [4] which led to the exclusion zones which were depicted in Figure 1.The plots for the FSPL and ITM diffraction/troposcatter loss are depicted in Figure 5, where the red (blue) curve indicates the free-space (ITM) losses versus the separation distances. It is noteworthy that the two models predict very close values for the losses in the LoS region, approximately 50 km, whereas this loss sharply elevates in the NLoS region, where ITM model is valid. As we mentioned, the border of the LoS region occurs at 49.5 km where the FSPL applies in accordance with the red curve which shows the amount by which radar signals are attenuated in dB as a function of their travelled distance. The maximum amount of FSPL is about 139 dB. Then, the radar signals are in the NLoS region and undergo the scatter loss which is the first sharp rise that we observe as the blue line, which reveals a drastic propagation loss that is accrued in this region. The, at roughly 90 km away from the ship, the radar signals suffer from the troposcatter loss which has a less increase as opposed to its preceding scatter loss.

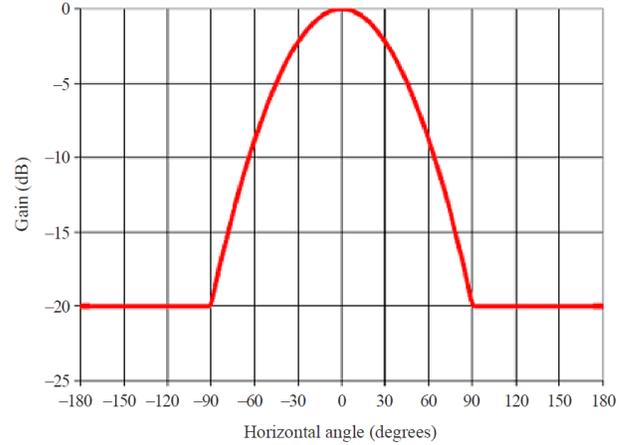

**Figure 4: Composite antenna pattern the eNBs for 3-sector cells from equation (4).**

It is noteworthy that in our simulations the distance of the radar from the cellular system is 50, 100, 150, and 200 km (Table 2). For the 50 km case, the radar is mainly in the LoS region and only 500 m is contained in the NLoS region which amounts to a maximum 155 dB loss. On the other hand, for 100, 150, and 200 km cases, respectively 50.5, 100.5, and 150.5 km are within the NLoS regions where the radar signals suffer both diffraction and propagation losses with an approximate maximum path loss of correspondingly 188, 193, and 200 dB.

Table 4: ITM Parameters (adopted from [4]).

| Parameters | Value |
|---|---|
| Operation Mode | APM |
| LTE/Radar Antenna Height | 25/50 m |
| Dielectric Constant | 15 |
| Conductivity | 0.005 S/m |
| Refractivity | 301 N-units |
| Climate | Continental Temperate |
| Variability Mode | Single Message |

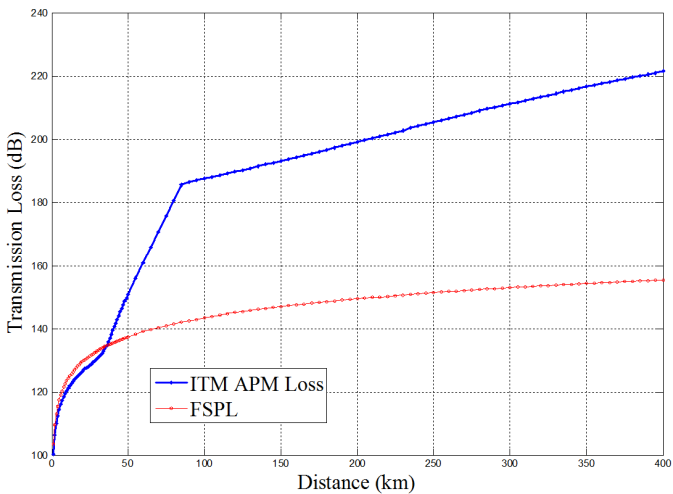

**Figure 5: LoS FSPL in red and NLoS ITM loss in blue represent the extent to which the radar signal degrades vs. the traveled distance. In the LoS/NLoS region (before/after 50 km in our simulations) the FSPL/ITM propagation losses represent the attenuation for the radar signal before reaching the eNBs.**

### III. SIMULATION RESULTS

We set the simulation time to 5 seconds, during which the the impact of the radiation of the radar with parameters in section II-A onto the eNBs of the LTE cellular system with parameters in section II-B, is investigated. The radar is cochannel with the LTE system and rotates 360 deg in azimuth in 445 beam positions where the radar sojourns for the dwell time 4.5 ms and sends 9 pulses 78 µs wide and 83 dBm through its 45 dBi antenna to the eNBs in the beam position based on equation (1). For instance, at the LTE system, the width of the radiation spans 1.5, 3.0, 4.5, and 6.0 km when the radar is respectively 50, 100, 150, and 200 km away from the LTE system. As such, all the eNBs covered by the radiation width would suffer from the radar pulses which are amplified by the transmitter and antenna and undermined by the propagation losses explained in section II-c. We investigate the interference from the radar for every Transmission Time Interval (TTI) [6] of the LTE, which is 71.4 µs.

The layout for the LTE system is depicted in Figure 6. Here the 7 red circles indicate 7 eNBs and the gray "+" symbols are the UEs which are uniformly distributed around the eNBs. There are 21 cells (7 sites and 3 cell per site) and 10 UEs per cell, which amounts to 210 UEs which move in a uniform direction at 3kmph. The area spans 1600 m × 1600 m, and the sector antenna pattern were depicted already in Figure 4.

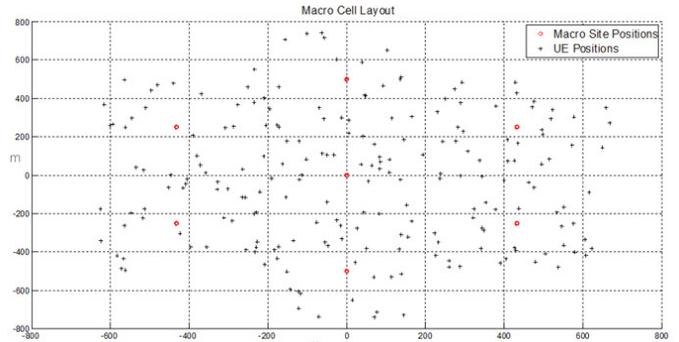

**Figure 6: LTE hexagonal macro cell layout with 7 eNBs and d 210 UEs, uniformly distributed in the area.**

First, we plot the SINR of an LTE eNB versus LTE symbol and subcarrier indices in Figure 7 where we observe an SINR drop due to the radar pulse affecting LTE symbols on the uplink during the simulation time. Interestingly, even when the radar is present, the SINR recovers back to its normal baseline situation until the next pulse hits the same region. We assume that the radar pulses hit the same geographic region, i.e. the same eNBs are affected, during the antenna dwell time (that is the same beam position). This is realistic as the amount of relocation for a large ship during PRIs is negligible.

Because the radar pulse is assumed to be centered in the LTE band, most of the pulse energy is concentrated around subcarrier 300 (in the middle of the LTE band which has 600 subcarriers [6]). Also, being at 78 us wide, the radar pulse slightly exceeds the duration of the LTE symbol (71.4 us). Thus, most of the energy is concentrated in symbol 1 and 8, with some remaining pulse energy also present in symbol 2 and 9. This is promising as only certain symbols of an LTE sub-frame are affected by the radar signal. It is worth mentioning that between symbols 1 and 8 there are 7 symbols which are each 71.4 µs. That is to say the time between these symbols, at which we observe the SINR drops due to the radar pulses in Figure 7, is 7 × 71.4 µs ≈ 0.5 ms which equals the radar PRI in accordance with table 2.

Next, we look at the mean throughput of the UEs affected by the radiating radar in Figure 8. Here, the brown bar represents the baseline where the radar is absent. On the other hand, the approaching radar decreases the throughput of the UEs. As we can observe from the green, orange, and light blue bars, when radar is 100, 150, and 200 km away from the LTE system, there is only a slight UE throughput decline whereas the throughput loss is slightly more pronounced for a radar only 50 km away from the LTE system. As it is expected, the further away the radar, the higher the mean UE throughput as interference becomes less pronounced due to the diffraction loss caused by the ITM in the NLoS region. We emphasize that the propagation path losses for the interference scenarios is based on the FSPL and ITM simulations whose plots were depicted in Figure 5.

It is noteworthy that Figure 8 represents relative values due to privacy considerations of Nokia. However, as we can observe from this figure, LTE is capable of operating within the NTIA assigned exclusion zones. In fact, the amount of LTE throughput loss for a radar as close as 50 km is about 20% whereas the loss for radars further away is negligible.

Moreover, Figure 8 does not precisely show the slight throughput losses incurred when the radar is 100, 150, and 200 km away vis-à-vis the baseline. In order to reveal the throughput degradation for the aforesaid distances more accurately, we resort to the cumulative distribution functions (CDFs) of the germane throughput losses, depicted in Figure 9. Here, the blue curve represents the baseline, i.e. the situation where no radar is present in the vicinity of the LTE system. However, as we can observe from the cyan, green, and black curves, the radar at distances of correspondingly 200, 150, and 100 km away from the LTE system generates some throughput loss. On ther other hand, the throughput degradation is more pronounced when radar is 50 km away from the LTE system (red curve).

It is to be noted that the abscissa values are omitted due to confidential considerations from Nokia. Albeit, to illustrate we draw a dashed line to intersect the CDFs. As we can see, when the radar is 50 km away, more than 50% of the UEs have a throughput less than the corresponding values obtained by the intersection of the abscissa and the dashed line. In contrast, less than 30% of the UEs degrade to the same throughput when the radar is 100, 150, and 200 km away from the shore.

Last but not least, it is worth mentioning that the results presented here are based on non-idealized (practical) modulation and coding rate selection algorithms for the uplink LTE scheduler [9]. Base on the results, we see that when a cochannel radar interference is present, care must be taken in deploying LTE eNBs within the coastal regions in the 3.5 GHz band.

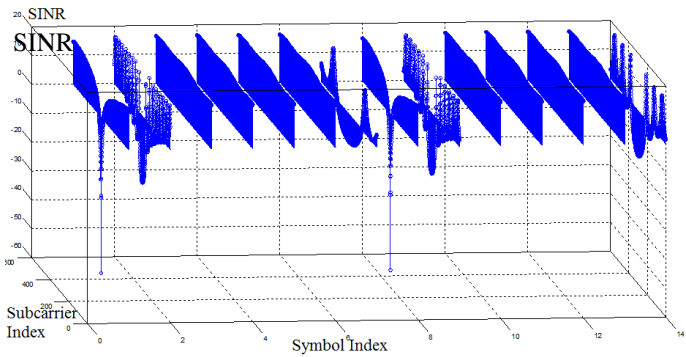

**Figure 7: SINR per symbol for the LTE eNB radiated by the radar interference reducing SINR significantly.**

On the other hand, even for a radar as close as 50 km, the throughput loss is not dramatic. This distance is far less than the exclusion zones proposed by NTIA [4] which were as large a 577 km inland. Therefore, the possibility of producing narrower exclusion zones to allow for leveraging the government-held spectrum in the 3.5 GHz band for mobile broadband within the coastal regions of the US is highly recommended.

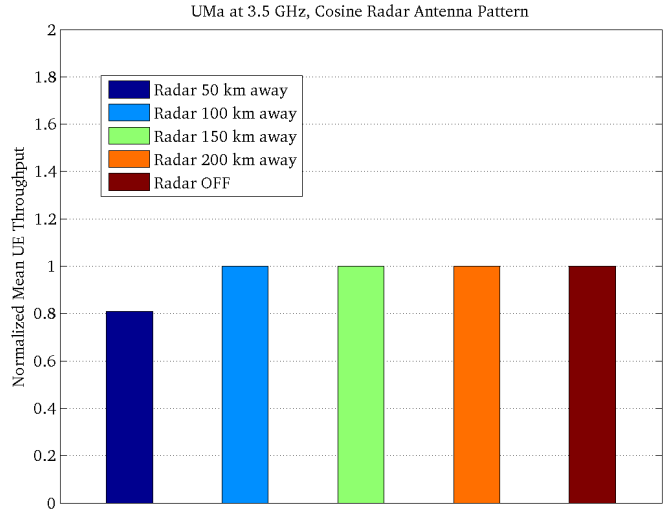

**Figure 8: Mean UE throughput at different radar distances to the LTE system.**

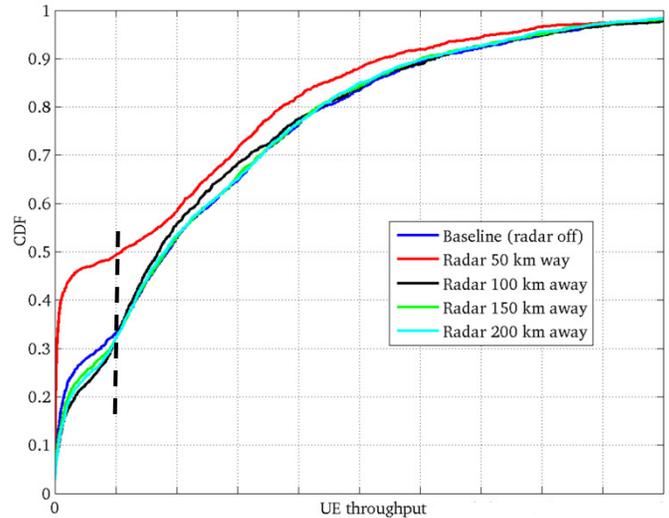

**Figure 9: CDF of the UE throughput at various distances of the radar to the LTE system.**

### IV. CONCLUSION

In this paper, we studied the impact of shipborne S-band shipborne radar systems that are cochannel with and in the vicinity of a cellular 3.5 GHz LTE systems. We leveraged the LTE standard to simulate LTE uplink at a system level by simulating the details for the LTE protocol compliant to the 3GPP standard. Furthermore, we simulated a radar system with parameters according to the NTIA report [4] as a bench mark in order to simulate similar radar systems. In addition, we considered radar antenna rotation and pattern in the simulations.

Similarly to the NTIA report which had led to the exclusion zones [4], we simulated free-space and ITM pathlosses conducive to modeling the LoS and diffraction/troposcatter losses according to which the radar signals were attenuated to obtain the relevant signal levels for the radar pulses once they arrived at the LTE system. In the simulations, we assessed the radar impact by observing the SINR for symbol and subcarrier indices. We observed that the presence of the radar reflects SINR plummets for the eNBs during the times pulses hit the eNBs. However, LTE was able to recover during the time between radar pulses, which was promising as only certain LTE symbols were affected.

Furthermore, we looked at the UE mean throughputs losses when a radar interference occurs. Contrasting the baseline with interference scenarios at various distances between the radar and LTE system, we realized that even though the radar interference is clearly visible by degrading the UE throughputs in the uplink direction, the throughput loss is tolerable even with radar deployed only 50 km away from the LTE system.

In view of the simulation results of the current article, the authors suggest a significant reduction in the NTIA-proposed exclusion zones. Such a reduction can constructively enable cellular providers to share the 3.5 GHz band with the radar systems and expand their services to the coastline metropolitan areas. Such an action not only ensues conspicuous revenues for the mobile network providers, but it also motivates fulfilling the spectrum sharing with government incumbents in the other bands as well.

REFERENCES

[1] Nokia Solutions and Networks White Paper – 'Enhance mobile networks to deliver 1000 times more capacity by 2020' http://nsn.com/sites/default/files/document/tv2020_1000x_capacity_wp.pdf.

[2] Executive Office of the President, President's Council of Advisors on Science and Technology (PCAST), "Realizing the Full Potential of Government-Held Spectrum to Spur Economic Growth", 2012

[3] Federal Communications Commission, "Proposal to Create a Citizen's Broadband Service in the 3550-3650 MHz band", FCC Docket No. 12-354, 2012.

[4] "An Assessment of the Near-Term Viability of Accommodating Wireless Broadband Systems in the 1675-1710 MHz, 1755-1780 MHz, 3500-3650 MHz, 4200-4220 MHz and 4380-4400 MHz Bands" NTIA, U.S. Dept. of Commerce, Nov. 2010

[5] J. Andrews, A. Ghosh, R. Muhamed, "Fundamenytals of WiMAX: Understanding Broadband Wireless Netwroking", Prentice Hall Communications Engineering and Emerging Technologies Series, 2007.

[6] A. Ghosh, R. Ratasuk, "Essentials of LTE and LTE-A", The Cambridge Wireless Essentials Series, September 26, 2011

[7] S. Wilson, T. Fischetto, "Coastline Population Trends in the United States: 1960 to 2008", U.S. Dept. of Commerce, 2010.

[8] G. Hufford, A. Longley, W. Kissick, "A Guide to the Use of the ITS Irregular Terrain Model in the Area Prediction Mode", U.S. Dept. of Commerce, 1982.

[9] 3GPP TR 36.814 V9.0.0 (2010-03), "Further advancements for E-UTRA physical layer aspects", Release 9, 2010. Measuring of Heterogeneous Wireless and Wired Networks, 2012.

[10] M. Cotton, R. Dalke, "Spectrum Occupancy Measurements of the 3550-3650 Megahertz Maritime Radar Band Near San Diego, California", NTIA, U.S. Dept. of Commerce, 2014.

[11] A. Lackpour, M. Luddy, and J. Winters, "Overview of Interference Mitigation Techniques between WiMAX Networks and Ground based Radar," in Proc. 20th Annual Wireless and Optical Comms. Conf., 2011.

[12] A. Khawar, A. Abdelhadi, C. Clancy, "Spectrum Sharing between S-band Radar and LTE Cellular System: A Spatial Approach", in IEEE Dynamic Spectrum Access Networks (DYSPAN), 2014.

[13] Ghorbanzadeh, M., Abdelhadi, A., Clancy, C.," A Utility Proportional Fairness Resource Allocation in Spectrally Radar-Coexistent Cellular Networks", MILCOM 2014.

[14] F. Sanders, J. Carrol, G. Sanders, R. Sole, " Effects of Radar Interference on LTE Base Station Receiver Performance", NTIA, U.S. Dept. of Commerce, 2013.

[15] M. Richards, J. Scheer, W. Holm, "Principles of Modern Radar", SciTech Publishing, 2010.

[16] "Guidelines for Evaluation of Radio Interface Technologies for IMT-Advanced", ITU-R M.2135-1, Dec. 2009.